\documentclass[
    ,final            
  ]
  {aipproc}

\layoutstyle{8x11single}

\newcommand{\MET}       {$E_T\hspace{-2.4ex}/\hspace{1.2ex}$}

\newcommand{\GeVcc}     {\ \mathrm{GeV/c^2}}

\newcommand{\GeV}       {\ \mathrm{GeV}}
\newcommand{\MeVcc}     {\ \mathrm{MeV/c^2}}
\newcommand{\invpb}     {~\mathrm{pb}^{-1}}
\newcommand{\invfb}     {~\mathrm{fb}^{-1}}

\newcommand{\mathMET}       {E_T\hspace{-2.4ex}/\hspace{1.2ex}}

\newcommand{\mathgl}        {\tilde{g}}

\newcommand{\neutralino} {$\tilde \chi_1^0$}
\newcommand{\mathneutralino} {\tilde \chi_1^0}
\newcommand{\mathneutralinoll} {\tilde \chi_2^0}
\newcommand{\mathchargino} {\tilde \chi_1^\pm}

\begin{document}

\title{Supersymmetry Searches at the Tevatron}


\author{Xavier Portell\footnote{Speaker, on behalf of the CDF and D\O\ 
Collaborations.}}{
  address={IFAE, Barcelona, Spain \\
           portell@fnal.gov}
}

\begin{abstract}
CDF and D\O\ detectors have already collected $1.3\invfb$ of data delivered by the Tevatron collider at 1.96~TeV center-of-mass energy. We present here the various analyses that are currently testing the possibility of a supersymmetric extension of the Standard Model. No evidence for such processes have been found in luminosities that range from $300$ to $800\invpb$ and different limits on the different supersymmetric models are set. Constraints coming from indirect searches are also presented.
\end{abstract}

\maketitle


\section{Introduction}

 Supersymmetry (SUSY)~\cite{susy} is one of the most appealing theoretical frameworks extending the Standard Model (SM). Among other benefits, it solves the hierarchy problem and favors the unification of forces at a Grand Unification scale. It is based on a new symmetry between bosons and fermions. As a result, for every known elementary particle a superpartner is predicted to exist differing by half a unit of spin and having masses at the TeV scale. To protect leptonic and baryonic number violations, a new quantum number is introduced, R-parity ($R_p$). When $R_p$ is conserved, supersymmetric particles are pair-produced and there is a lightest SUSY particle (LSP) which is stable and escapes detection. If the lightest neutralino (\neutralino) is the LSP, it is a leading candidate for astronomically observed cold dark matter and gives missing transverse energy (\MET) signatures in the detector. When $R_p$ is broken, events without large \MET\ will be observed in the detector.

We present the main SUSY results from the Tevatron experiments. $R_p$ is assumed to be conserved throughout the text except when violation is explicitly stated. These analyses and other that could not be covered here can be found at the web pages of CDF~\cite{CDFweb} and D\O~\cite{D0web}.

\section{Squark and Gluino searches}

\subsection{General searches using \MET\ and Jets}
\begin{figure}
  \includegraphics[height=.3\textheight]{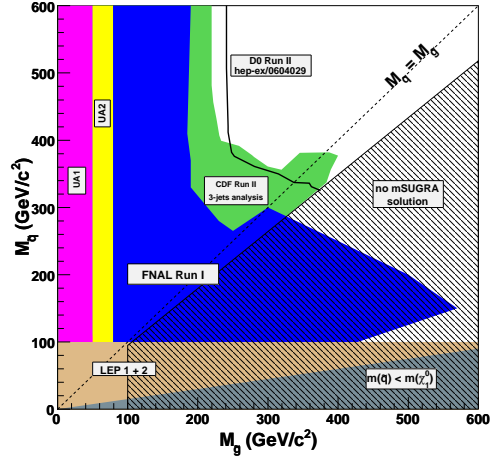}
  \caption{CDF and D\O\ observed exclusion limits at 95\% C.L. for squark and gluino searches. Previous results are also included. The dashed zone is the non-mSUGRA solution.}
\label{fig:sqgl_limits}
\end{figure}

We report on the search for the superpartners of quarks and gluons, which can be pair produced at the Tevatron via strong interactions. The ultimate decay of these particles leads to event topologies characterized by the presence of multiple jets, from the cascade decays of heavy sparticles, and large \MET\ in the final state, from the LSP (\neutralino). CDF (D\O~\cite{D0sqgl}) studies a mSUGRA scenario~\cite{msugra} with $\mu<0$, $A_0=0$ and $\tan\beta=5$ ($\tan\beta=3$). The different SM background contributions include QCD multijet production, with \MET\ coming from fluctuations in the jet energies measured in the calorimeter, and boson plus jets, diboson and top production, with \MET\ produced by neutrino presence in the final states. Whenever possible, SM backgrounds are normalized to next-to-leading order (NLO) predictions. QCD multijet processes have been simulated using Monte Carlo and then normalized to the data (CDF) or extrapolated from the data (D\O).
CDF and D\O\ analyses are optimized for different regions in the gluino-squark mass plane depending on the gluino mass (CDF) or the number of jets (D\O). Both experiments use similar luminosity samples: $371\invpb$ (CDF) $310\invpb$ (D\O).
Data agree with expected SM yields and different exclusion limits at 95\% C.L. are set (see Figure~\ref{fig:sqgl_limits})\footnote{CDF and D\O\ collaborations compute the limits following different strategies.}. In the region where gluino and squark masses are similar, masses below $387\GeVcc$ are excluded. In any case, squark masses are excluded below $325\GeVcc$ and gluino masses are excluded below $241\GeVcc$. 

\subsection{Search for $3_{\rm rd}$ Generation Squarks}
In mSUGRA, the mixing between the left- and right-handed squarks is large for the third generation. Thus, the lightest supersymmetric squarks might be either the stop or the sbottom and they could be the first hint of SUSY at colliders. 

\subsubsection{Sbottom}

\begin{figure}
\includegraphics[height=.3\textheight]{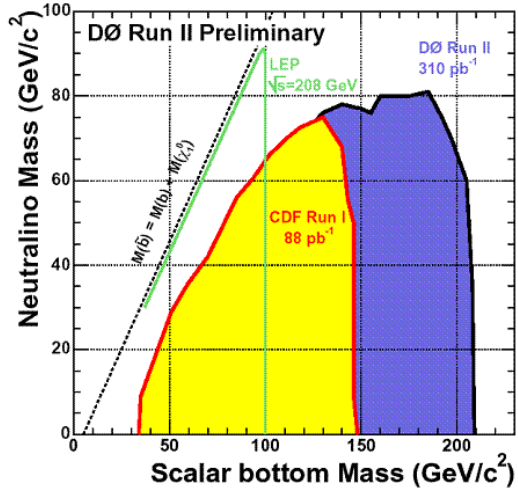}
\caption{D\O\ observed exclusion limits at 95\% C.L. for sbottom searches.}
\label{fig:sbottom}
\end{figure}

The two main decay modes for sbottom pair production at the Tevatron are:
\begin{itemize}
\item $q\bar q\to \bar{\tilde{b}}\tilde b\to \mathneutralino\bar bb\mathneutralino$
\item $q\bar q\to \mathgl\mathgl\to \bar b \tilde{b} \bar{\tilde{b}} b\to \mathneutralino \mathneutralino b \bar b b \bar b$
\end{itemize}

\noindent where, in all the cases, a mSUGRA scenario is chosen such that the branching ratio of $\tilde{b}\to b\mathneutralino$ is close to 100\%. The D\O\ collaboration studied the first of these channels requiring at least two jets and \MET. In order to keep the acceptance high, only one of the jets is required to be $b$-tagged. In addition, to reduce the SM background contribution which mainly comes from weak bosons and top production, different thresholds for \MET\ and transverse momentum of the jets are applied as a function of the sbottom mass. No deviations from standard model expectations were found. Figure~\ref{fig:sbottom} shows the observed limit. The different possible sbottom masses are kinematically constrained by the neutralino mass, $M_{\tilde b} > M_b + M_{\mathneutralino}$. Sbottom masses up to $200\GeVcc$ are excluded.

\subsubsection{Stop}

The stop presents different decay channels depending on the mass. The light stop decay into $c\mathneutralino$ is the most explored signature at the Tevatron but recent D\O\ studies gain insight into a heavier stop, which goes through a three body decay via a virtual chargino:
$$\tilde{t_1}\bar{\tilde{t}_1}\to b\bar b b\mathchargino \to b\bar b l^\pm\tilde\nu\to l^\pm\nu\mathneutralino\ .$$

\begin{figure}
\includegraphics[height=.3\textheight]{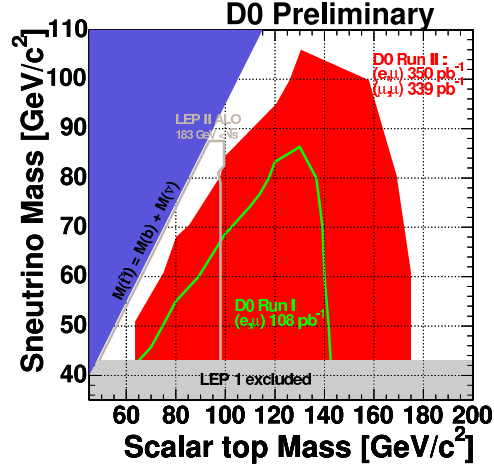} 
\caption{D\O\ observed exclusion limits at 95\% C.L. for stop searches.}
\label{fig:stop}
\end{figure}

Theoretically, chargino decays into $W\mathneutralino$ are also possible but they are disfavored for the higher mass of the $W$ boson. Therefore, the decay of $\mathchargino$ into $\tilde\nu$ and a lepton is assumed to be 100\% and the slepton parameters have been set to obtain equal branching fraction for the different lepton flavors. Two different final states are considered depending on the lepton flavor: $e\mu$ or $\mu\mu$. In addition, two different relative values between the stop and the scalar neutrino masses are also studied. Around $350\invpb$ are considered in the analyses and some topological cuts together with a cut on $\mathMET>15$~GeV are implemented. Since no deviation from SM expectations are observed, new exclusion limits are extracted as shown in Figure~\ref{fig:stop}. Stop masses up to approximately the SM top mass are excluded.

\section{Charginos and Neutralinos: multileptons and \MET}
We present searches for associated production of the lightest chargino ($\mathchargino$) and the second lightest neutralino ($\mathneutralinoll$) in mSUGRA models. These processes present a very clean signature of three leptons and \MET. One of the leptons comes from the $\mathchargino \to l^\pm\nu\mathneutralino$ decay, and the other two come from $\mathneutralinoll \to l^\pm l^\mp\mathneutralino$ decay. The \MET\ comes from the presence of the \neutralino 's and neutrinos in the final states. Different analyses are performed searching for three leptons where the third one has low transverse momentum. To increase acceptance, two leptons with an extra track and two like-sign leptons coming from different decay chains are also considered. This allows to include searches with taus in the final state whose production is enhanced for large $\tan\beta$ values.

\begin{figure}
  \includegraphics[height=.3\textheight]{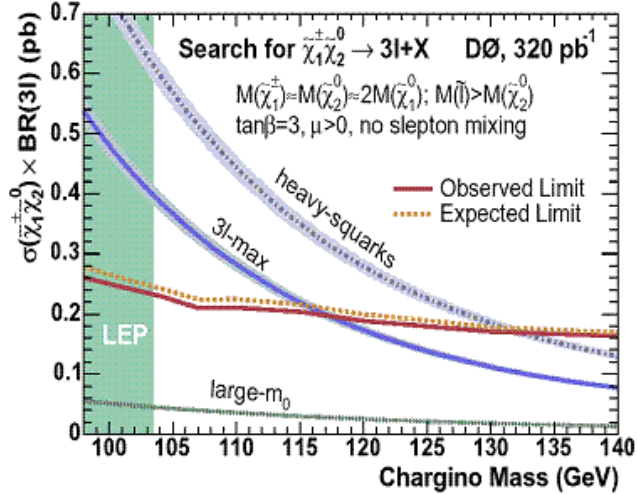}
  \caption{D\O\ exclusion limits for chargino-neutralino analyses.}
\label{fig:trileptons}
\end{figure}

Background processes that can fake the multilepton plus \MET\ signature involve Drell-Yan plus a misidentified jet or photon conversion, and dibosons. CDF results for the different channels are summarized in Table~\ref{tab:trileptons}. No significant excess is observed and limits are set on the $\mathchargino$ mass by combining the different results. In a low $\tan\beta$ scenario and considering slepton mass degeneration, CDF sets a limit of $M_{\tilde{\mathchargino}}>127\GeVcc$. D\O\ also studied different decay channels and found no evidence of associated chargino-neutralino production. Figure~\ref{fig:trileptons} shows D\O\ limits in three different scenarios, considering low $\tan\beta$ and no slepton mixing~\cite{D0trilept}. For the first time, CDF and D\O\ limits improve LEP results for this choice of parameters. 

\begin{table}
\begin{tabular}{lrrrr}
\hline
\tablehead{1}{c}{b}{Mode}  & \tablehead{1}{c}{b}{Luminosity} & 
\tablehead{1}{c}{b}{SM expectation} & 
\tablehead{1}{c}{b}{Observed}\\
\hline
$ee+l$       & 346 & $0.17\pm 0.05$ & 0\\
$e\mu +l$    & 700 & $0.78\pm 0.11$ & 0\\
$\mu\mu + l$ & 745 & $0.64\pm 0.18$ & 1\\
$\mu\mu +l$\tablenote{Low $p_T$ trigger} & 312 & $0.13\pm 0.03$ & 0\\
$ee + {\rm track^*}$ & 607 & $0.49\pm 0.10$ & 1\\
$l^\pm l^\pm$  & 704 & $6.8\pm 1.0$ & 9 \\
\hline

\end{tabular}
\caption{CDF results on searches for $\mathchargino \mathneutralinoll$ production.}
\label{tab:trileptons}
\end{table}

\section{GMSB: diphotons}
In Gauge-Mediated Supersymmetry Breaking (GMSB) models, the LSP is the superpartner of the graviton (gravitino). The neutralino is the next to the lightest supersymmetric particle and decays into a gravitino and one photon. Thus, associated production of chargino-neutralino produces two photons and two gravitinos which escape detection, leaving a \MET\ signature. D\O\ has recently studied this channel with $760\invpb$ of data requiring two photons with transverse energies of $25\GeV$ and $\mathMET>45\GeV$. The SM expected background after cuts constitutes $2.1\pm0.7$ events from $e\gamma$ and QCD multijets. One event is observed, being compatible with the SM. A limit is set excluding charginos below $220\GeVcc$ and neutralino below $120\GeVcc$, which greatly extends previous results~\cite{D0GMSB}.

\section{R-Parity violation (RPV)}

\subsection{Multilepton events in \neutralino\ decay}
In the most general SUSY lagrangian there exist some terms that explicitly violate lepton and baryonic numbers. These terms receive strong constraints from other non-collider experiments like the measurement of the lifetime of the proton. The part of the lagrangian that violates the leptonic number is written as
\begin{equation}
R_p\hspace{-2.4ex}/\hspace{1.2ex}=\lambda_{ijk}L_iL_j\bar E_k \; ,
\end{equation}

\noindent where $ijk$ indicate the lepton families involved. If $R_p$ is not conserved, searches for these processes allow $R_p$ conservation in all the decay vertices except in the last decay, where the LSP must decay to SM particles. Very clean and distinctive signatures involving four leptons and \MET, coming from neutrinos, are expected. CDF and D\O\ are investigating $\lambda_{121}$ term, which involves signatures such as $eell$ ($l=e,\mu$), and $\lambda_{122}$ term, which produces $\mu\mu ll$ signatures. D\O\ is also studying $\lambda_{133}$ term, involving third lepton family, which is expected to be dominant at high $\tan\beta$.

The events have been selected with at least three leptons. No excesses from the SM expectations are found and limits are extracted for the different $\lambda$ couplings in specific mSUGRA scenarios. In particular, D\O\ excludes charginos (neutralino) below $231\GeVcc$ ($119\GeVcc$) for $\lambda_{121}$ processes with $\mu>0$, $m_0=1$~TeV and $\tan\beta=5$. The limits obtained for $\lambda_{133}$ at small $m_0$ and $\tan\beta\approx 20$, which exclude the chargino (neutralino) below $217\GeVcc$ ($115\GeVcc$), are competitive with the other studies for $\lambda_{121}$ and $\lambda_{122}$.

\subsection{RPV stop}
\begin{figure}
  \includegraphics[height=.3\textheight]{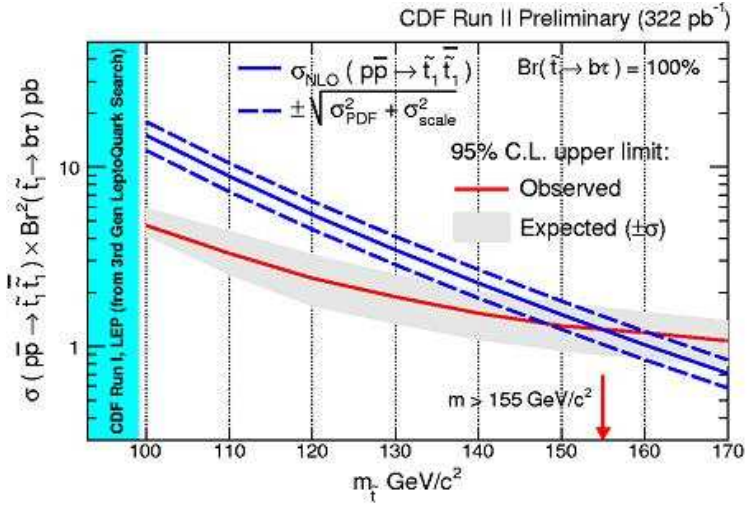}
  \caption{CDF observed and expected limits for the stop searches in RPV scenarios.}
  \label{fig:RPVstop}
\end{figure}

CDF has also performed a search for $\tilde{t_1}\bar{\tilde{t_1}}$ production in an $R_p$ violating scenario in which the stop decays to $b\tau$ at 100\% of the time, assuming the $\lambda'_{333}$ coupling non-zero. Events are selected with two $b$-jets, one hadronic $\tau$ decay and one semi-leptonic $\tau$ decay. The efficiency for $\tau$-tagging is estimated to be around 56\% from $Z\to\tau\tau$ studies. With a sample of $322\invpb$, two events are found, one with an electron in the final state and one with a muon. This result is in agreement with SM expectations of $2.2^{+0.46}_{-0.22}$. A 95\% C.L. limit is set on the stop mass of $155\GeVcc$ ($151\GeVcc$ if a more conservative approach, reducing the theoretical uncertainties by one standard deviation, is considered) as shown in Figure~\ref{fig:RPVstop}. This result can also be interpreted as a limit on the third generation leptoquark ($LQ_3$), which become identical to this process when assuming $BR(LQ_3\to\tau b)=1$.

\section{Indirect Searches: $B_s\to\mu\mu$}
This indirect search is a powerful tool to search for physics beyond the SM. This decay is heavily suppressed within the SM, which predicts $BR(B_s\to\mu\mu)=(3.5\pm 0.9)\times 10^{-9}$. However, SUSY can enhance it through loops (proportional to $\tan^6\beta$) or via  $R_p$ violating processes. Recently, CDF updated the studies on the $B_s\to\mu\mu$ rare decay with $760\invpb$ of data. In order to implement a precise measurement of this rare process, CDF use $B^+\to J/\psi K^+$ as a control sample for normalization purposes. In addition, a very precise dimuon mass resolution of $\sigma(\mu\mu)\approx 0.23\MeVcc$ is required to distinguish $B_d$ and $B_s$ states. One event is found after event selection, which is compatible with SM expectations. The new limit on this process is computed to be $BR(B_s\to\mu\mu)<1.0\times 10^{-7}$. This limit improves the latest published result by a factor of two~\cite{CDFbsmumu}.

\section{Conclusions}
The results presented here, with luminosities between 300 and $800\invpb$, improve previous limits on searches for Supersymmetry at the Tevatron and at LEP2. The Tevatron experiments have collected more than $1.3\invfb$ of data since the beginning of Run II and will continue in operation until 2009. The new data that will be collected by CDF and D\O\ detectors will provide an opportunity for new physics discovery before the LHC starts.


\begin{theacknowledgments}
  I would like to thank the organizers for the kind invitation
and their great hospitality during the conference.
\end{theacknowledgments}



\bibliographystyle{aipproc}   

\bibliography{sample}


\end{document}